\documentclass[oribibl, orivec]{llncs}
\usepackage{times}

\usepackage{amsmath}
\usepackage{amssymb}
\usepackage{url}
\usepackage{tikz} 
\usepackage{listings}
\usepackage{amsmath}
\usepackage{amssymb}
\usepackage[ruled,vlined]{algorithm2e}

\newcommand{\semb}{[ \! [}
\newcommand{\seme}{] \! ]}
\newcommand{\comment}[1]{}

\def\norm#1{\mbox{$\| #1 \|_1$}}

\def\norminfty#1{\mbox{$\| #1 \|_\infty$}}

\def\transpose#1{{}^t \! #1}

\def\middle{\operatorname{mid}}			% mod/dev as operators
\def\dev{\operatorname{dev}}
\def\vect#1{\mathchoice{\mbox{\boldmath$\displaystyle#1$}}	% define \vect as a bold face notation (LNCS style of \vec)
{\mbox{\boldmath$\textstyle#1$}}
{\mbox{\boldmath$\scriptstyle#1$}}
{\mbox{\boldmath$\scriptscriptstyle#1$}}}

\def\cpform{constrained affine set}
\def\new{\mbox{new }}

\def\ua{\uparrow_\circ}
\def\da{\downarrow_\circ}
\def\ra{\rightarrow}
\def\bbr{{\mathbb R}}

\def\R{{\mathbb R}}

\newcommand {\N}{{\rm{I\!N}}}

\newlength{\parskipLightVerbatim}
{\obeyspaces\gdef {\ }}
\def\TCode{\def\[{[}\def\]{]}\catcode`\[=1 \catcode`\]=2 \catcode`\{=12
\catcode`\}=12 \catcode`\&=12 \catcode`\#=12
\catcode`\~=12 \catcode`\_=12 \catcode`\^=12 \obeyspaces \tt}

\begin{document}
\def\bbr{{\Bbb R}}
\def\ua{\uparrow_\circ}
\def\da{\downarrow_\circ}
\def\ra{\rightarrow}

\def\uan{\uparrow_n}
\def\dan{\downarrow_n}
\def\uam{\uparrow_-}
\def\dam{\downarrow_-}
\def\uap{\uparrow_+}
\def\dap{\downarrow_+}

% boldface letters for intervals
\def\bfm#1{\protect{\makebox{\boldmath $#1$}}}
\def\a {\bfm{a}}
\def\b {\bfm{b}}
\def\c {\bfm{c}}
\def\d {\bfm{d}}
\def\e {\bfm{e}}
\def\f {\bfm{f}}
\def\g {\bfm{g}}
\def\h {\bfm{h}}
\def\ii {\bfm{i}}       % \i is already 'i without dot'
\def\j {\bfm{j}}
\def\k {\bfm{k}}
\def\l {\bfm{l}}
\def\m {\bfm{m}}
\def\n {\bfm{n}}
\def\o {\bfm{o}}
\def\p {\bfm{p}}
\def\q {\bfm{q}}
\def\r {\bfm{r}}
\def\s {\bfm{s}}
\def\t {\bfm{t}}
\def\u {\bfm{u}}
\def\vv {\bfm{v}}       % \v is already 'check' 
\def\w {\bfm{w}}
\def\x {\bfm{x}}
\def\y {\bfm{y}}
\def\z {\bfm{z}}
\def\B {\bfm{B}}
\def\C {\bfm{C}}
\def\DD{\bfm{D}}        % \D is already \displaystyle
\def\E {\bfm{E}}
\def\G {\bfm{G}}
\def\H {\bfm{H}}
\def\J {\bfm{J}}
\def\L {\bfm{L}}
\def\M {\bfm{M}}
\def\N {\bfm{N}}
\def\O {\bfm{O}}
\def\P {\bfm{P}}
\def\Q {\bfm{Q}}
\def\S {\bfm{S}}
\def\T {\bfm{T}}
\def\U {\bfm{U}}
\def\V {\bfm{V}}
\def\W {\bfm{W}}
\def\X {\bfm{X}}
\def\Y {\bfm{Y}}
\def\Z {\bfm{Z}}

\title{A Logical Product Approach to Zonotope Intersection}

\author{Khalil Ghorbal, Eric Goubault, Sylvie Putot}
\institute{%Commissariat \`a l'\'Energie Atomique\\ 
Laboratory for the Modelling and Analysis of Interacting Systems \\
CEA, LIST, Bo\^ite 94, Gif-sur-Yvette, F-91191 France.\\
\email{firstname.lastname@cea.fr}}

\maketitle

\begin{abstract}
We define and study a new abstract domain which is a fine-grained combination of 
zonotopes with polyhedric domains such as the interval, octagon, linear templates or
polyhedron domain. While abstract transfer functions are still 
rather inexpensive and accurate even for interpreting non-linear
computations, we are able to also interpret tests (i.e. intersections) efficiently.
This fixes a known drawback of zonotopic methods, as used for reachability analysis for
hybrid systems as well as for invariant generation in abstract interpretation: 
intersection of zonotopes are not always zonotopes, and there is not even a best
zonotopic over-approximation of the intersection. 
We describe some examples and an implementation of our method in the APRON library, and 
discuss some further interesting combinations of zonotopes with non-linear or non-convex
domains such as quadratic templates and maxplus polyhedra.
\end{abstract}

\section{Introduction}
\label{introduction}

Zonotopic abstractions are known to give fast and accurate over-appro\-ximations
in invariant synthesis for static analysis of programs, as introduced
by the authors \cite{NSAD05,SAS06,CAV09}, as well as in reachability analysis of 
hybrid systems \cite{gir-05}.
The main reason for this is that the interpretation of linear
assignments is exact and done in linear time in terms of the ``complexity''
of the zonotopes, and non-linear
expressions are dynamically linearized in a rather inexpensive way, unlike
for most of other sub-polyhedric domains (zones \cite{zones}, linear templates \cite{lineartemplates}, even
polyhedra \cite{polyhedra}). But unions, at the exception of recent work \cite{arxiv2},
and more particularly intersections \cite{LeGuernic08} are not canonical operations, and 
are generally computed using approximate and costly methods, contrarily
to the other domains we mentioned. We present in this
article a way to combine the best of the two worlds: by constructing 
a form of logical product \cite{logicalproduct}
of zonotopes with any of these sub-polyhedric domains, 
we still get accurate and inexpensive methods to deal with the interpretation
of linear and non-linear assignments, while intersections in particular,
come clean thanks to the sub-polyhedric component of the domain.

Consider for instance the following program (loosely based on 
non-linear interpolation methods in e.g. embedded systems), which will be our running example: 
\begin{lstlisting}[frame=single]
real x = [0,10];  
real y = x*x - x;
if (y >= 0) y = x/10; /* (x=0 or x >= 1) and y in [0,1] */
else y = x*x+2;       /* (x>0 and x<1) and y in [2,3] */
\end{lstlisting}
\vskip -.3cm
As indicated in the comments of the program, the \verb!if! branch is taken when we have $x=0$ or $x \geq 1$, so that $y$ 
at the end of the program, is always in $[0,3]$. Although this program looks quite simple, it is difficult 
to analyze and the invariants found for $y$ at the end of the program by classical domains\footnote{The experiments 
were carried out using the domains interfaced within APRON \cite{APRON}.} are disappointing: intervals, octagons, 
polyhedra, or zonotopes without constraint all find a range of value for $y$ larger or equal to $[0,102]$: 
even those who interpret quite accurately non-linear operations are not able to derive a constraint on $x$ 
from the constraint on $y$. 
Whereas by the method proposed here, a logical product of zonotopes with intervals, in its APRON implementation, 
we find the much better range $[0,9.72]$ (instead of the exact result $[0,3]$). 

\paragraph{Contents of the paper}
We first introduce in Section \ref{sect_recap} affine sets, a zonotopic abstract domain 
for abstract interpretation, that abstracts input/output relations in a program. We then introduce the
problem of computing intersections in Section \ref{sect_intuition}: starting with the running example, 
we define constrained affine sets as the combination of zonotopes with polyhedric domains and show 
they are well suited for the interpretation of tests. We then generalize the order on affine sets 
to constrained affine sets and define monotonic abstract transfer functions for arithmetic operators,
that over-approximate the concrete semantics.
Section \ref{sect_ordertheoretic} completes the definition of this new abstract domain: 
starting with the easier ``one-variable'' problem, we then give an algorithm for computing a join operator.   
We demonstrate the interest of the domain by describing in Section \ref{experiments} the results on some examples, 
based on an implementation of our method in the library APRON. 
We conclude by a discussion of future work, including some further interesting combinations of zonotopes with non-linear or non-convex
domains such as quadratic templates and maxplus polyhedra.

\paragraph{Related work}

In \cite{subpolyhedra}, the authors propose an approach based on a reduced product \cite{systematic},
to get more tractable and efficient methods for deriving sub-polyhedric invariants. 
But, still, the reduction algorithm of \cite{subpolyhedra} is fairly expensive, and this domain
also suffers from the drawbacks of polyhedra, in the sense that it is not well suited for efficiently
and precisely deriving invariants for non-linear computations.
Logical products in abstract interpretation are defined 
in \cite{logicalproduct}. The authors use the Nelson-Oppen combination method for logical theories, in
the convex case, to get polynomial time abstractions on a much finer (than classical reduced products) 
combination of two abstract
domains. As explained in Section \ref{sect_order}, this approach does not directly carry over our case,
because the theories we want to
combine do not satisfy all the hypotheses of \cite{logicalproduct}. We thus choose in this paper a direct
approach to the logical product %, that allows efficient over-approximations of the logical product 
of zonotopes with other classical abstract domains. 

%%%%%%%%%%%%%%%%%%%%%%%%%%%%%%%%%%%%%%%%%%%%%%%%%%%%%%%%%%%%%%%%%%%%%%%%%
\section{Affine sets: main definitions and properties}
\label{sect_recap}
\subsection{Affine arithmetic and zonotopes}
Affine arithmetic is an extension of interval arithmetic on affine forms, first introduced in \cite{com-sto-93-aa}, 
that takes into account affine correlations between variables. 
An {\em affine form} is a formal sum over a set of {\em noise symbols} 
$\varepsilon_i$ \[ \hat x \stackrel{def}{=} \alpha^x_0 + \sum_{i=1}^n \alpha^x_i \varepsilon_i,\]
with $\alpha^x_i \in \R$ for all $i$. %Let $\RAff$ denote the set of such affine forms. 
Each noise symbol $\varepsilon_i$ stands for an independent component of the total
uncertainty on the quantity $\hat x$, its value is unknown but bounded in [-1,1]; 
the corresponding coefficient $\alpha^x_i$ is a known real value, which gives 
the magnitude of that component. The same noise symbol can be 
shared by several quantities, indicating correlations among them. These noise 
symbols can not only model uncertainty in data or parameters,
but also uncertainty coming from computation.
The semantics of affine operations is straightforward, non affine operations are linearized and introduce 
a new noise symbol: we refer the reader to \cite{SAS06,arxiv1} for more details.

In what follows, we introduce matrix notations to handle tuples of affine forms.
We note ${\cal M}(n,p)$ the space of matrices with $n$ lines and $p$ columns of real coefficients.
A tuple of affine forms expressing the set of values taken by $p$ variables over $n$ noise symbols 
$\varepsilon_i, \; 1 \leq i \leq n$, can be represented by a matrix $A \in {\cal M}(n+1,p)$. 
We formally define the zonotopic concretization of such tuples by~: 
\begin{definition}\label{concretizations}
Let a tuple of affine forms with $p$ variables over $n$ noise symbols, defined by a matrix $A \in {\cal M}(n+1,p)$. 
Its concretization is the zonotope
$$\gamma(A)=\left\{\transpose{A} \transpose{e}  \mid e \in R^{n+1}, e_0=1, \norminfty{e}\leq 1\right\} \subseteq \R^p \enspace . $$
\end{definition}
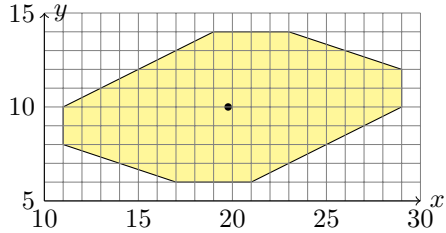
\begin{tikzpicture}[scale=0.25, left, information text/.style={inner sep=1ex}] 
\filldraw[fill=yellow!50!white] (29,10) -- (29,12) -- (23,14) -- (19,14) -- (11,10) -- (11,8) -- (17,6) -- (21,6) -- (29,10); 
\node (C) [fill=black,inner sep=1pt,shape=circle] at (20,10)  {};
\draw[step=1.cm,gray,very thin] (10,5) grid (30,15); 
\draw[->] (10,5) -- (30,5) node[right]{$x$} ;
\draw[->] (10,5) -- (10,15) node[right]{$y$};
\foreach \x/\xtext in {10, 15, 20, 25, 30} 
\draw (\x cm,5 cm)  node[below] {$\xtext$};
\foreach \y/\ytext in {5, 10, 15} 
\draw (10 cm,\y cm) node[left] {$\ytext$};

\draw[xshift=32.5cm, yshift=9.5cm]
  node[right,text width=5.5cm,information text]
  {
For example, for $n=4$ and $p=2$, 
the yellow zonotope is the concretisation of the affine set $X=(\hat{x},\hat{y})$,
with 
$\hat x = 20-4 \varepsilon_1 +2 \varepsilon_3 + 3 \varepsilon_4$, % \label{affex1}
$\hat y = 10-2\varepsilon_1 + \varepsilon_2 - \varepsilon_4$, %\label{affex2}
and
$\transpose{A}=\left(\begin{array}{ccccc}
20 & -4 & 0 & 2 & 3 \\
10 & -2 & 1 & 0 & -1
\end{array}\right).
$
  };
\end{tikzpicture} 
\subsection{An ordered structure: affine sets}
\label{perturbedforms-section}
In order to construct an ordered structure preserving abstract input/output relations \cite{arxiv2}, 
we now define affine sets $X$ as Minkowski sums of a {\em central} zonotope, $\gamma(C^X)$ 
and of a {\em perturbation} zonotope centered on $0$, $\gamma(P^X)$. Central zonotopes depend on 
central noise symbols $\varepsilon_i$, whose interpretation is fixed once and for all in 
the whole program: they represent the uncertainty on input values to the program, with
which we want to keep as many relations as possible. Perturbation zonotopes depend on 
perturbation symbols $\eta_j$ which are created along the interpretation of the program and
represent the uncertainty of values due to the control-flow abstraction, for instance while computing
the join of two abstract values. 

\begin{definition}
We define an affine set $X$ by the pair of matrices \\ $(C^X, P^X) \in {\cal M}(n+1,p) \times {\cal M}(m,p)$. 
The affine form $\pi_k(X)=c^X_{0k}+\sum_{i=1}^n c^X_{ik} \varepsilon_i +\sum_{j=1}^m p^X_{jk} \eta_j$,
where the  $\varepsilon_i$ are the central noise symbols and the $\eta_j$ the perturbation or union 
noise symbols, describes the $k$th variable of $X$.
\end{definition}

We define an order on affine sets \cite{CAV09,arxiv2} which is slightly more strict than
concretization inclusion: it formalizes the fact that the central symbols have a specific
interpretation as parametrizing the initial values of input arguments to the analyzed program:
\begin{definition}\label{preorder-def}
Let $X=(C^X, P^X)$, $Y=(C^Y, P^Y)$ be two affine sets $\,$ in ${\cal M}(n+1,p) \times {\cal M}(m,p)$. 
We say that $X\leq Y$ iff 
\[ \forall u \in \R^p, \; \norm{(C^Y-C^X)u} \leq \norm{P^Yu} - \norm{P^Xu} \enspace . \]
\end{definition}
It expresses that the norm of the difference $(C^Y-C^X)u$ for all $u \in \R^p$
is less than what the perturbation terms $P^X$ and $P^Y$ allow, that is the difference
of the norms of $P^Y u$ with $P^X u$.

The binary relation $\leq$ of Definition \ref{preorder-def} is a preorder, that
we identify in the sequel with the partial order, quotient of this preorder by the 
equivalence relation\footnote{It can be characterized by $C^X=C^Y$ and same concretizations for $P^X$ and $P^Y$.} 
$X \sim Y$ iff by definition $X \leq Y$ and $Y \leq X$.
Note also that this partial order is decidable, with a complexity bounded by a polynomial in $p$ and an
exponential in $n+m$. In practise, see \cite{arxiv2}, we do not need to use this costly 
general decision procedure. 

\section{{\em Constrained} affine sets for intersection}
\label{sect_intuition}

As discussed in \cite{arxiv1}, we can define an efficient join operator of two affine sets, but 
no satisfying meet operator. We now describe a logical product approach to interpret tests by adding
constraints in an abstract domain over the noise symbols $\varepsilon_i$. 
We first present the interpretation of tests, before other abstract transfer function, 
since this is the point that motivates the introduction of this new domain.

Take for example the running example. Before the test \verb!if (y >= 0)!, we have $x=5 + 5 \varepsilon_1$ and, 
using the multiplication on affine forms as defined in Section \ref{sect_arithmetic}, 
$y = 32.5 + 45 \varepsilon_1 + 12.5 \varepsilon_2 $, with $\varepsilon_1 \in [-1,1]$ and $\varepsilon_2 \in [-1,1]$.
In the abstract value associated with the \verb!true! branch is added the abstraction of the constraint $32.5 + 45 \varepsilon_1 + 12.5 \varepsilon_2 \geq 0$. Here, in intervals, it does not generate any additional 
constraint but it would in a relational domain. 
In the \verb!false! branch, the abstraction of $32.5 + 45 \varepsilon_1 + 12.5 \varepsilon_2 \leq 0$ is added, 
which generates constraint $\varepsilon_1 \leq -0.444$: we thus infer that $x$ is bounded by $[0,2.77]$ in this branch.  

We now introduce the logical product of the domain ${\cal A}_1$ of Section 
\ref{sect_recap} 
with any lattice, 
$({\cal A}_2,\leq_{2},\cup_{2},\cap_{2})$, 
used to abstract the values of
the noise symbols $\varepsilon_i$ and $\eta_j$. 
Formally, supposing that we have $n+1$ noise symbols $\varepsilon_i$
and $m$ noise symbols $\eta_j$ as in Section \ref{perturbedforms-section}, 
we are given a concretization function:
$\gamma_{2}: {\cal A}_2 \rightarrow {\cal P}(\{1\}\times\R^n\times \R^m)$
and pseudo-inverse $\alpha_{2}$.
We now define constrained affine sets:

\begin{definition}
A constrained affine set $U$ is a pair $U=(X,\Phi^X)$ where $X=(C^X,P^X)$
is an affine set, and $\Phi^X$ is an element of ${\cal A}_2$. 
Equivalently, we write $U=(C^X,P^X,\Phi^X)$.
\end{definition}

Classical abstractions of ``constraints'' on the $\varepsilon_i$ we will be using throughout this
text are ${\cal A}$ consisting of products of $1+n+m$ intervals (with the first one always being equal
to 1), zones, octagons, and polyhedra (in the hyperplane $\varepsilon_0=1$).

\subsection{Interpretation of tests}
\label{testsec}

\subsubsection{Equality tests on variables}
We first consider the case of the interpretation of equality test of two variables within 
an abstract state. 
Let us begin by a motivating example, which will make clear what the general interpretation
of Definition \ref{lb-eq-def} should be.

\begin{example}
Consider, with an interval domain for the noise symbols, $Z= \semb x_1==x_2 \seme X$ where
\label{ex_intersect}
$$
\left\lbrace \begin{array}{l} \Phi^X = 1 \times [-1,1]\times[-1,1]\times[-1,1]\\ 
\hat{x}_1^X = 4 + \varepsilon_1 + \varepsilon_2 + \eta_1, \quad \gamma(\hat{x}_1)=[1,7] \\ 
\hat{x}_2^X = - \varepsilon_1 + 3 \varepsilon_2,  \quad \gamma(\hat{x}_2)=[-4,4] \\
\end{array} \right.
$$
We look for $\hat z={\hat x}_1={\hat x}_2$, with $\hat z=z_0 + z_1 \varepsilon_1 + z_2 \varepsilon_2 + z_3 \eta_1$. 
Using $\hat{x}_1-\hat{x}_2=0$, i.e. 
\begin{equation}
\label{eq-constr}
4+2\varepsilon_1-2\varepsilon_2+ \eta_1=0,
\end{equation} and substituting $\eta_1$ in $\hat z-{\hat x}_1=0$, 
we deduce $z_0=4z_3$, $z_1=2z_3-1$, $z_2=-2z_3+3$. The abstraction in intervals of constraint (\ref{eq-constr}) 
yields tighter bounds on the noise symbols: $\Phi^Z=1 \times [-1,-0.5]\times[0.5,1]\times[-1,0]$.
We now look for $z_3$ that minimizes the width of the concretization of $z$, that is $0.5|2z_3-1|+0.5|3-2z_3|+|z_3|$. A straightforward $O((m+n)^2)$ method to solve the problem evaluates this expression for $z_3$ successively equal to $0$, $0.5$ and $1.5$: the minimum is reached for $z_3=0.5$. We then have
$$
\left\lbrace \begin{array}{l} \Phi^Z = 1 \times [-1,-0.5]\times[0.5,1]\times[-1,0]\\ 
\hat{x}_1^Z = \hat{x}_2^Z (= \hat{z}) = 2 + 2\varepsilon_2 + 0.5 \eta_1, \quad \gamma(\hat{x}^Z_1)=\gamma(\hat{x}^Z_2)=[2.5,4]
 \end{array} \right.
$$
Note that the concretization $\gamma(\hat{x}_1^Z)=\gamma(\hat{x}_2^Z)$ is not only better than the intersection of the concretizations
$\gamma(\hat{x}_1^X)$ and $\gamma(\hat{x}_2^X)$ which is $[1,4]$, but also better than the intersection of the concretization of 
affine forms $(\hat{x}_1^X)$ and $(\hat{x}_2^X)$ for noise symbols in $\Phi^Z$. %It is not always the case 
Note that there is not always a unique solution minimizing the width of the concretization.
\end{example}
In the general case, noting in bold letters the interval concretization of noise symbols:
\begin{definition}
\label{lb-eq-def}
Let $X = (C^X,P^X,\Phi^X)$ a constrained affine set with $(C^X,P^X) \in 
{\cal M}(n+1,p) \times {\cal M}(m,p)$. We define $Z= \semb x_j==x_i \seme X$ by:\\
- $\Phi^Z = \Phi^X \bigcap \alpha_{2}\left(\left\{(\varepsilon_1,\ldots,\varepsilon_n,\eta_1,\ldots,\eta_m) \mid  
(c_{0j}^X-c_{0i}^X) + \sum_{r=1}^{n}(c_{rj}^X-c_{ri}^X) \varepsilon_r + \right.\right.$ \\ $\left. \left. \sum_{r=1}^{m}(p_{rj}^X-p_{ri}^X) \eta_r= 0 \right\}\right)$, \\
- $c^Z_{rl}=c^X_{rl}, \forall r \in \{0,\ldots,n\},$ and $\forall l \in \{1,\ldots,p\}, l \neq i, j$,\\
- $p^Z_{rl}=p^X_{rl}, \forall r \in \{1,\ldots,m\}$ and $\forall l \in \{1,\ldots,p\}, l \neq i, j$.\\
Let $k$ such that $c_{kj}^X-c_{ki}^X \neq 0$, we define
\begin{equation}
\label{eq-cond-c}
c^Z_{li}=c^Z_{lj}= c^X_{li}+\frac{c^X_{lj}-c^X_{li}}{c^X_{kj}-c^X_{ki}}(c^Z_{ki}-c^X_{kj}) \; \forall l \in \{0, \ldots,n\}, \; l \neq k,
\end{equation}
\begin{equation}
\label{eq-cond-p}
p^Z_{li}=p^Z_{lj}=p^X_{li}+\frac{p^X_{lj}-p^X_{li}}{c^X_{kj}-c^X_{ki}}(c^Z_{ki}-c^X_{kj}) \; \forall l \in \{1, \ldots,m\},
\end{equation}
with $c^Z_{ki}$ that minimizes $\sum_{l=1}^{n} |c^Z_{li}| \dev(\vect{\varepsilon_l^Z}) + \sum_{l=1}^{m} | p^Z_{li} | \dev(\vect{\eta_l^Z})$.\\
\end{definition}
This expresses that the abstraction of the constraint on the noise symbols induced by the test is added to the domain of constraints, and the exact
constraint is used to define an affine form $z$ satisfying $z=x^Z_j=x^Z_i$, and such that  $\gamma(z)$ is minimal.
Indeed, let k such that $c_{kj}^X-c_{ki}^X \neq 0$, then $x_j == x_i$ allows to express $\varepsilon_k$ as
\begin{equation}
\label{cond-noise} 
\varepsilon_k = c_{0j}^X-c_{0i}^X + \sum_{1 \leq l \leq n, l \neq k} \frac{(c_{lj}^X-c_{li}^X)}{c_{ki}^X-c_{kj}^X} \varepsilon_l +  
\sum_{1 \leq l \leq m} \frac{(p_{lj}^X-p_{li}^X)}{c_{ki}^X-c_{kj}^X} \eta_l \enspace . 
\end{equation}
 We now look for $\pi_i(Z)=\pi_j(Z)$ equal to $\pi_i(X)$ and 
$\pi_j(X)$ under condition (\ref{cond-noise}) on the noise symbols. Substituting $\varepsilon_k$ in for example $\pi_i(Z)=\pi_i(X)$, we 
can express, for all $l$, $c^Z_{li}$ and $p^Z_{li}$ as functions of $c^Z_{ki}$ and get possibly an infinite number of solutions defined by 
(\ref{eq-cond-c}) and (\ref{eq-cond-p}) that are all equivalent when (\ref{cond-noise}) holds.
When condition (\ref{cond-noise}) will be abstracted in a noise symbols abstract domain such as intervals, these abstract solutions 
will no longer be equivalent, we choose the one that minimizes the width of 
$\gamma(\pi_i(Z))$ which is given by $\sum_{l=1}^{n} |c^Z_{li}| \dev(\vect{\varepsilon_l^Z}) + \sum_{l=1}^{m} |p^Z_{li}| \dev(\vect{\eta_l^Z})$.
This sum is of the form $\sum_{l=1}^{m+n} |a_l + b_l c^Z_{ki}| $, with known constants $a_l$ and $b_l$. The minimization problem 
can be efficiently solved in $O((m+n)log(m+n))$ time, $m+n$ being the number of noise symbols appearing in the expressions of $x_i$ and $x_j$, 
by noting that the minimum is reached for $c^Z_{ki}=-\frac{a_{l_0}}{b_{l_0}}$ for a $l_0 \in \{1,\ldots,m+n\}$. When it is 
reached for two indexes $l_p$ and $l_q$, it is reached for all $c^Z_{ki}$ in $\bigl[ -\frac{a_{l_p}}{b_{l_p}},-\frac{a_{l_p}}{b_{l_p}}\bigr]$,
but we choose one of the bounds of this intervals because it corresponds to the substitution in 
$x_i^Z$ of one of the noise symbols, and is in the interest for the interpretation of tests on expressions.

\subsubsection{Equality tests on expressions}
Now, in the case of an equality test between arithmetic expressions, new constraints on the noise symbols can be added, 
corresponding to the equality of the two expressions interpreted as affine forms. We also choose new affine forms for 
variables appearing in the equality test:
let $X = (C^X,P^X,\Phi^X)$ a constrained affine set with $(C^X,P^X) \in 
{\cal M}(n+1,p) \times {\cal M}(m,p)$. We define $Z= \semb exp1==exp2 \seme X$ by:
$Y_1= \semb x_{p+1} = exp1 \seme \semb x_{p+2} = exp2 \seme X$ using the semantics for 
arithmetic operations, as defined in section \ref{sect_arithmetic}, 
then $Y_2=\semb x_{p+1}==x_{p+2} \seme Y_1$. Noting that one of the noise symbols appearing in the constraint 
introduced by the equality test, does not appear in $x_{p+1}^{Y_2}=x_{p+2}^{Y_2}$
as computed by Definition \ref{lb-eq-def}, using this constraint we  substitute this noise symbol in the other variables in $Y_2$. 
We then eliminate the added variables $x_{p+1}$ and $x_{p+2}$ to obtain $Z$, in which 
$exp1==exp2$ is thus algebraically satisfied.

\begin{example}
\label{ex_intersect2}
Consider $Z= \semb x_1 + x_2 ==x_3 \seme X$ where
$$
\left\lbrace \begin{array}{l} \Phi^X = 1 \times [-1,1]\times[-1,1]\times[-1,1]\\ 
\hat{x}_1^X = 2 + \varepsilon_1, \quad \gamma(\hat{x}_1)=[1,3] \\ 
\hat{x}_2^X = 2 + \varepsilon_2 + \eta_1, \quad \gamma(\hat{x}_2)=[0,4]\\
\hat{x}_3^X = - \varepsilon_1 + 3 \varepsilon_2,  \quad \gamma(\hat{x}_3)=[-4,4]
\end{array} \right.
$$
We first compute $x_4 := x_1+x_2$ in affine arithmetic: here, problem $x_4 == x_3$ is then the test we solved in 
example \ref{ex_intersect}. The abstraction in intervals of  constraint  (\ref{eq-constr})
 yields $\Phi^Z=1 \times [-1,-0.5]\times[0.5,1]\times[-1,0]$, 
and an affine form $x_3^Z$ optimal in the sense of the width of its concretization, $x_3^Z=2 + 2\varepsilon_2 + 0.5 \eta_1$.
Now, $\hat{x}_1^X+\hat{x}_2^X=\hat{x}_3^Z$ is satisfied when constraint (\ref{eq-constr}) holds exactly, but not in its 
interval abstraction $\Phi^Z$. But substituting $\varepsilon_1$ which does not appear in $x_3^Z$ by $-2+\varepsilon_2-0.5\eta_1$ 
in $\hat{x}_1^X$ and $\hat{x}_2^X$, we obtain forms $\hat{x}_1^Z$ and $\hat{x}_2^Z$ that satisfy $x_1 + x_2 ==x_3$
in the abstract domain:
$$
\left\lbrace \begin{array}{l} \Phi^Z = 1 \times [-1,-0.5]\times[0.5,1]\times[-1,0]\\ 
\hat{x}_1^Z =  \varepsilon_2 - 0.5 \eta_1, \quad \gamma(\hat{x}_1)=[0.5,1.5] \\ 
\hat{x}_2^Z =  2 + \varepsilon_2 + \eta_1, \quad \gamma(\hat{x}_2)=[1.5,3]\\
\hat{x}_3^Z =  2 + 2\varepsilon_2 + 0.5 \eta_1, \quad \gamma(\hat{x}^Z_1)=\gamma(\hat{x}^Z_2)=[2.5,4]\\
 \end{array} \right.
$$

\end{example}

\subsubsection{Inequality tests}
In the case of inequality tests, we only add constraints on noise symbols, for example for strict inequality:
\begin{definition}
\label{lb-ineq-def}
Let $X = (C^X,P^X,\Phi^X)$ a constrained affine set with $(C^X,P^X) \in 
{\cal M}(n+1,p) \times {\cal M}(m,p)$. We define $Z= \semb exp1 < exp2 \seme X$ by $Z= (C^X,P^X,\Phi^Z)$:\\
$\Phi^Z = \Phi^X \bigcap \alpha_{2}\left(\left\{(\varepsilon_1,\ldots,\varepsilon_n,\eta_1,\ldots,\eta_m) \mid  
(c_{0p+2}^Y-c_{0p+1}^Y) + \right.\right.$ \\
$\left.\left. \sum_{k=1}^{n}(c_{kp+2}^Y-c_{kp+1}^Y) \varepsilon_k + \sum_{k=1}^{m}(p_{kp+2}^Y-p_{kp+1}^Y) \eta_k < 0 \right\}\right),$\\ 
where $Y = \semb x_{p+1}= exp1 \seme \semb x_{p+2} = exp2 \seme X$.
\end{definition}

\subsection{Order relation}

\label{sect_order}
In a standard reduced product \cite{systematic} of ${\cal A}_1$ with ${\cal A}_2$, the order relation 
would naturally be based on the component-wise ordering. But in such products, we cannot possibly reduce the
abstract values so that to gain as much collaboration as needed
between ${\cal A}_1$ and ${\cal A}_2$ for giving formal grounds
to the reasoning of Example \ref{ex_intersect} for instance. 
What we really need is to combine the {\em logical theories}
of affine sets, $Th({\cal A}_1)$\footnote{Signature $\Sigma_{{\cal A}_1}$ comprises equality, addition, multiplication
by real numbers and real numbers.}
with 
the one of quantifier-free linear arithmetic~\cite{Manna} over the reals, 
, $Th({\cal A}_2)$\footnote{Signature $\Sigma_{{\cal A}_2}$ 
comprises $\Sigma_{{\cal A}_1}$ plus inequality and negation}, 
including all the domains we have in mind in this paper (intervals, zones, octagons,
linear and non-linear templates, polyhedra). Look back at Example \ref{ex_intersect}: we
found a solution to the constraint $x_1==x_2$ via a fine-grained interaction between the two
theories $Th({\cal A}_1)$ and $Th({\cal A}_2)$. 
Unfortunately, the methods of 
\cite{logicalproduct} are not directly applicable; in particular ${\cal A}_1$ is not 
naturally expressible as a logical lattice - it is not even a lattice in general. Also, 
the signatures $\Sigma_{{\cal A}_1}$ and $\Sigma_{{\cal A}_2}$ share common symbols, which is not allowed in the approach
of \cite{logicalproduct}.

In order to compute the abstract transfer functions in the logical product $Th({\cal A}_1) \cup Th({\cal A}_2)$,
we first define an order relation on the product domain ${\cal A}_1
\times {\cal A}_2$, that allows a fine interaction between the two domains. 
First, $X=(C^X,P^X,\Phi^X) \leq Y=(C^Y,P^Y,\Phi^Y)$ should imply that $\Phi^X \leq_{2} \Phi^Y$, i.e. the range of values
that noise symbols can take in form $X$ is smaller than for $Y$. Then,
we mean to adapt Definition \ref{preorder-def} for noise symbols no longer defined in $[-1,1]$
as in the unconstrained case, but in the range of values $\Phi^X$ common to $X$ and $Y$.
Noting that:
$$\norm{C^X u}=\sup_{\epsilon_i \in [-1,1]} |\langle \epsilon, C^X u\rangle |,$$
where $\langle .,. \rangle$ is the standard scalar product of vectors
in $\R^{n+1}$, we set:
\begin{definition}
\label{constrainedpreorder}
Let $X$ and $Y$ be two constrained affine sets. We say that $X \leq Y$ iff
$\Phi^X \leq_2 \Phi^Y$ and, for all $t \in \R^p$,
$$\sup_{\epsilon_i,\eta_j \in \gamma_2(\Phi^X)}  | \langle (C^Y-C^X) t,\varepsilon\rangle |
\leq \sup_{\epsilon_i,\eta_j \in \gamma_2(\Phi^Y)}  | \langle P^Y t,\eta\rangle | -
\sup_{\epsilon_i,\eta_j \in \gamma_2(\Phi^X)}  | \langle P^X t,\eta\rangle |  \enspace .$$
\end{definition}

The binary relation defined in Definition \ref{constrainedpreorder} is a preorder on constrained
affine sets which coincides with Definition \ref{preorder-def}
in the ``unconstrained'' case when $\Phi^X=\Phi^Y=\{1\}\times [-1,1]^{n+m}$. 
We use in the sequel its quotient by its equivalence relation, i.e.
the partial order generated by it.

\begin{definition}
Let $X$ be a constrained affine set. Its concretization in ${\cal P}(\R^p)$ is 
$$\gamma(X)=\left\{\transpose{C^X} \epsilon+\transpose{P^X} \eta \mid \epsilon_i, \eta_j \in \gamma_2(\Phi^X) \right\} \enspace .$$
\end{definition}

For $\Phi^X$ such that $\gamma_2(\Phi^X)=\{1\}\times [-1,1]^{n+m}$, this
is equivalent to the concretization of the affine set $(C^X, P^X)$ as
defined in Section \ref{perturbedforms-section}.
As for affine sets \cite{arxiv2}, the order relation of Definition
\ref{constrainedpreorder} is stronger than the geometric order:
if $X \leq Y$ then $\gamma(X) \subseteq \gamma(Y)$.
This allows for expressing functional dependencies between the input
and current values of each variables as discussed in \cite{arxiv2}.

Note that $\gamma$ is in general computable. %: this point is important
In case ${\cal A}$ is a sub-polyhedric domain, such as intervals, zones,
octagons, linear templates and polyhedra, $\gamma$ can be computed
using any (guaranteed) solver for linear programs
such as LURUPA \cite{lurupa}, since computing
$\gamma$ involves two linear programs:
$$ \sup_{\epsilon_i, \eta_j \in \gamma(\Phi^X)} \transpose{C^X} \epsilon
+\transpose{P^X} \eta, \mbox{ and } \inf_{\epsilon_i, \eta_j \in \gamma(\Phi^X)} \transpose{C^X} \epsilon
+\transpose{P^X} \eta \enspace .$$

\subsection{Semantics of arithmetic operations}

\label{sect_arithmetic}

Operations are not different than the ones generally defined on zonotopes,
or on affine forms, see \cite{com-sto-93-aa,arxiv2}, the only difference is in
the multiplication where we use the constraints on $\epsilon_i$ and 
$\eta_j$ to derive bounds for the non-linear part.

We note $\semb \new \epsilon_{n+1} \seme_{{\cal A}_2} \Phi^X$ the creation of a new noise symbol $\epsilon_{n+1}$ with 
(concrete) values in $[-1,1]$.
We first define the assignment of a new variable $x_{p+1}$ with a range of value $[a,b]$:
\begin{definition}\label{assignment}
Let $X=(C^X,P^X,\Phi^X)$ be a \cpform $\,$ with $(C^X,P^X) \in 
{\cal M}(n+1,p) \times {\cal M}(m,p)$ %dimension $((n+1)\times p, m\times p$
and $a, b\in \R$. We define $Z= \semb x_{p+1}=[a,b] \seme X$
where $(C^Z,P^Z) \in {\cal M}(n+2,p+1) \times {\cal M}(m,p+1)$ %to be of dimension $((n+1)\times (p+1), m\times (p+1)$ 
with~:
$\Phi^Z=\semb \new \epsilon_{n+1} \seme_{{\cal A}_2} \Phi^X$,  % HIPHOP 
$C^Z=\left(\begin{array}{c|c}
& \frac{a+b}{2} \\
& 0 \\
C^X 
& \ldots \\
& 0\\
\hline
0 & \frac{\mid a-b \mid}{2}
\end{array}\right)$, 
$P^Z=\left(\begin{array}{c|c}
& 0 \\
P^X & \ldots \\
& 0 \\
\end{array}\right) \enspace .$
\end{definition}

We carry on by addition, or more precisely, the operation interpreting
the assignment $x_{p+1}:=x_i+x_j$ and adding new variable $x_{p+1}$ to the affine set:

\begin{definition}\label{addition}
Let $X=(C^X,P^X,\Phi^X)$ be a \cpform $\,$ where $(C^X,P^X)$ is
in ${\cal M}(n+1,p) \times {\cal M}(m,p)$. %of dimension $((n+1)\times p,m\times p)$. 
We define $Z=\semb x_{p+1}=x_i + x_j \seme X=(C^Z,P^Z,\Phi^Z)$ where
$(C^Z,P^Z)\in {\cal M}(n+1,p+1) \times {\cal M}(m,p+1)$ by 
$\Phi^Z=\Phi^X$ and \\
$$C^Z=\left(\begin{array}{c|c}
C^X & \begin{array}{c}
c^X_{0,i}+c^X_{0,j} \\
\ldots \\
c^X_{n,i}+c^X_{n,j} \\
\end{array}
\end{array}\right) \mbox{ and } \;
P^Z=\left(\begin{array}{c|c}
P^X & \begin{array}{c}
p^X_{1,i}+p^X_{1,j} \\
\ldots \\
p^X_{m,i}+p^X_{m,j} \\
\end{array}
\end{array}\right) \enspace .$$
\end{definition}

The following operation defines the multiplication of
variables $x_i$ and $x_j$, appending the result to 
the \cpform $\,X$. All polynomial assignments can be defined using this and
the previous operations. 

\begin{definition}\label{multiplication}
Let $X=(C^X,P^X,\Phi^X)$ be a \cpform $\,$ where $(C^X, P^X)$ is
in ${\cal M}(n+1,p) \times {\cal M}(m,p)$. 
We define $Z=(C^Z,P^Z,\Phi^Z)=\semb x_{p+1}=x_i \times x_j \seme X$
where $(C^Z,P^Z) \in {\cal M}(n+2,p+1) \times {\cal M}(m+1,p+1)$ by~:
\begin{itemize}
\item $\Phi^Z=\semb \new \epsilon_{n+1} \seme_{{\cal A}_2} \circ
\semb \new \eta_{m+1} \seme_{{\cal A}_2} \Phi^X$
\item $c^z_{l,k}=c^x_{l,k}$ and
$c^z_{n+1,k}=0$ for all $l=0,\ldots,n$ and
$k=1,\ldots,p$
\item Let $m_r$ (resp. $\mu_r$) be the $(r+1)$th coordinate (i.e. corresponding to $\varepsilon_r$)
of $\middle(\gamma(\Phi^X))$ (resp. of $\dev(\gamma(\Phi^X))$), 
where $\middle{}$ (resp. $\dev{}$) denotes the middle (resp. the radius)
of an 
interval, $q_l$ (resp. $\chi_l$) be
the $(l+n+1)$th coordinate (i.e. corresponding
to $\eta_l$)
of $\middle(\gamma(\Phi^X))$ (resp. of $\dev(\gamma(\Phi^X))$). 
Write $d_i^x=c_{0,i}^x+\sum_{1\leq r \leq n} c^x_{r,i}m_r
+\sum_{1\leq l \leq m} p^x_{l,i} q_l$:

$c^z_{0,p+1}=d^x_i d^x_j-\sum_{1 \leq r \leq n}
(d^x_i c^x_{r,j}+d^x_j c^x_{r,i})m_r -\sum_{1 \leq l \leq m} (d^x_i p^x_{l,j}+d^x_i p^x_{l,i})q_l
+\sum_{1 \leq r \leq n} \frac{1}{2} c^x_{r,i} c^x_{r,j} m_r^2+\sum_{1 \leq l \leq m} \frac{1}{2} p^x_{l,i} p^x_{l,j} \q_l^2$ 
\item $c^z_{l,p+1}=d^x_i c^x_{l,j}+c^x_{l,i} d^x_j$
for all $l=1,\ldots,n$
\item $c^z_{n+1,p+1}=\sum_{1 \leq r \leq n} \frac{1}{2} | c^x_{r,i} c^x_{r,j}
| \mu_r^2
+\sum_{1\leq r\neq l \leq n} | c^x_{r,i} 
c^y_{l,j} | \mu_r \mu_l$ 
\item $p^z_{l,k}= p^x_{l,k}$, $p^z_{m+1,k}=0$ and $p^z_{l,p+1}=0$, for
all $l=1,\ldots,m$ and $k=1,\ldots,p$
\item $p^z_{m+1,p+1}=\sum_{1 \leq l \leq m} | p^x_{l,i} p^x_{l,j} |
\chi_l^2$
$+\sum_{1 \leq r\neq l \leq m} | p^x_{r,i} p^x_{l,j} |
\chi_r \chi_l$
$+ \sum_{0 \leq r \leq n}^{1 \leq l \leq m} (| c^x_{r,i} p^x_{l,j} |
$
$+| p^x_{l,i} c^x_{r,j}|) \mu_r \chi_l$. 
\end{itemize}
\end{definition}
The correctness of this abstract semantics stems from the fact that
these operations are increasing functions over the set of constrained
affine sets.
For sub-polyhedric domains ${\cal A}_2$, $m_r$, $q_l$, $\mu_r$ and $\chi_l$ are easily computable, solving with a guaranteed
linear solver the four linear programming problems $\sup_{\epsilon,\eta \in \gamma(\Phi^X)} \epsilon_r$ (resp. $\inf$) and
$\sup_{\epsilon,\eta \in \gamma(\Phi^X)} \eta_l$ (resp. $\inf$) - for an interval domain for ${\cal A}_2$, no such computation 
is needed of course.

Getting back to the running example of Section \ref{introduction}, in
the \verb!false! branch of the \verb!if (y>=0)! test, we have to compute  $y=x*x+2$
with $x=5+5\varepsilon_1$ and $\varepsilon_1 \in [-1,-0.444]$. 
Using Definition \ref{multiplication} which takes advantage of the bounds on $\varepsilon_1$ to get a 
better bound on the non-linear part (typically not possible if we had constructed a reduced product), we get 
$y=14.93+13.9\varepsilon_1+0.96\varepsilon_3$ with $\varepsilon_3 \in [-1,1]$. 
This gives $\gamma(y)=[0.07,9.72]$, which is very precise since $\gamma(x)=[0,2.77]$, hence we should ideally find $\gamma(y)$ in 
$\gamma(x)*\gamma(x)+2=[2,9.72]$. Note that the multiplication given in Definition \ref{multiplication} and used here, is not the 
direct adaptation of the multiplication in the unconstrained case, that would give the much less accurate form
$y=41.97+50 \varepsilon_1+10.03\varepsilon_3$: the better formulation is obtained by choosing an affine form that is a linearization
of $x_i\times x_j$ non longer at $0$, but at the center of the range of the constrained noise symbols.

%%%%%%%%%%%%%%%%%%%%%%%%%%%%%%%%%%%%%%%%%%%%%%%%%%%%%%%%%%%%%%%%%%%%%%%%%
\section{Join operator on constrained affine sets}
\label{sect_ordertheoretic}

We first examine the easier case of finding a join operator for affine sets with just one variable, and 
${\cal A}_2$ being the lattice of intervals. 
We then use the characterisations we find in this case to give efficient formulas for a
precise (although over-approximated) join operator in the general case. We do not study here maximal lower bounds
of affine sets, although they are naturally linked to the interpretation of tests, Section \ref{testsec}, this is outside the scope of 
this paper.

\subsection{The one-dimensional case}
\label{onedim}
In dimension one, constrained affine sets are simply \emph{constrained affine forms}:
$$
\hat{a}=\Big(\hat{a}(\epsilon)=\alpha_0^a + \sum_{1}^n{\alpha_i^a\epsilon_i},\ \beta^a,\ \Phi^a \Big),
$$
where $\varepsilon = (\epsilon_1, \dots{}, \epsilon_n)^t$ belongs to $\Phi^a$, and $\beta^a$ is non negative.
We use the bold face notation, $\vect{\epsilon_i^a}$, to denote the interval concretization of $\epsilon_i$. %, the $\imath$\textsuperscript{ th} component of $\varepsilon$.
Let $\hat{a}$ and $\hat{b}$ be two constrained affine forms.
Then $\hat{a} \leq \hat{b}$ in the sense of Definition
\ref{constrainedpreorder} if and only if
$$
\left\lbrace \begin{array}{l} \Phi^a \subseteq \Phi^b  \\ 
\sup_{\varepsilon \in \Phi^a} 
|\hat{a}(\varepsilon) - \hat{b}(\varepsilon)| \leq \beta^b - \beta^a
\end{array} \right.
$$
In general, there is no least upper bound for two constrained affine forms, but rather, as already
noted in the unconstrained case \cite{arxiv1,arxiv2}, {\em minimal upper bounds}. %: $\hat{z}$ is a minimal upper bound (mub) of $\hat{x}$ and 
A sufficient conditions for $\hat{c}$ to be a minimal upper bound is to enforce a minimal concretization,
that is, $\gamma(\hat{c}) = \gamma(\hat{a}) \cup \gamma(\hat{b})$, and then minimize $\beta^c$ among upper bounds with this concretization.

Algorithm~\ref{algo:mub_itv_min} computes this particular mub in some cases (when the first {\tt return} branch is taken), and else an upper bound 
with minimal interval concretisation.
Let us introduce the following notion used in the algorithm: let $\vect{i}$ and $\vect{j}$ be two intervals; 
$\vect{i}$ and $\vect{j}$ are said to be
in generic position if $(\vect{i} \subseteq \vect{j}$ or $\vect{j} \subseteq \vect{i})$ imply $(\sup(\vect{i})=\sup(\vect{j})$ or $\inf(\vect{i})=\inf(\vect{j}))$. 
We say by extension that two affine forms are in generic position if their interval concretisations are in generic position. 
The join algorithm is similar to the formula in the unconstrained
case described in \cite{arxiv2} except we have to be cautious about the relative position of the ranges of noise symbols.

\begin{algorithm}
\DontPrintSemicolon
  \BlankLine
\If {$\hat{a}$ {\bf and} $\hat{b}$ are in generic position} {
  \lIf {$\middle(\gamma(\hat{b})) \leq \middle(\gamma(\hat{a}))$} {swap $\hat{a}$ and $\hat{b}.$\;}
  \For{$i \geq 1$}{
    $\alpha_i^c \longleftarrow 0$\;
    \If {$\vect{\epsilon_i^a}$ and $\vect{\epsilon_i^b}$ are in generic position}{
      \If {$\alpha_i^a \geq 0$ {\bf and} $\alpha_i^b \geq 0$} {
        \If {$\middle(\vect{\epsilon_i^a}) \leq \middle(\vect{\epsilon_i^a} \cup \vect{\epsilon_i^b})$ {\bf and} $\middle(\vect{\epsilon_i^b}) \geq \middle(\vect{\epsilon_i^a} \cup \vect{\epsilon_i^b})$} {
           $\alpha_i^c \longleftarrow \min(\alpha_i^a,\alpha_i^b)$\;
         }
      }
      \If {$\alpha_i^a \leq 0$ {\bf and} $\alpha_i^b \leq 0$} {
        \If {$\middle(\vect{\epsilon_i^a}) \geq \middle(\vect{\epsilon_i^a} \cup \vect{\epsilon_i^b})$ {\bf and} $\middle(\vect{\epsilon_i^b}) \leq \middle(\vect{\epsilon_i^a} \cup \vect{\epsilon_i^b})$} {
          $\alpha_i^c \longleftarrow \max(\alpha_i^a,\alpha_i^b)$\;
        }
      }
    }
    }
    \If {$0 \leq \sum_{i=1}^{n}\alpha_i^c(\middle(\vect{\epsilon_i^a} \cup \vect{\epsilon_i^b}) - \middle(\vect{\epsilon_i^a})) \leq \middle(\gamma(\hat{a}) \cup \gamma(\hat{b})) - \middle(\gamma(\hat{a}))$ {\bf and} $\middle(\gamma(\hat{a}) \cup \gamma(\hat{b})) - \middle(\gamma(\hat{b})) \leq \sum_{i=1}^{n}\alpha_i^c(\middle(\vect{\epsilon_i^a} \cup \vect{\epsilon_i^b}) - \middle(\vect{\epsilon_i^b})) \leq 0$} {
       $\beta^c \longleftarrow \dev(\gamma(\hat{a}) \cup \gamma(\hat{b})) - \sum_{i=1}^{n} |\alpha_i^c|\dev(\vect{\epsilon_i^a} \cup \vect{\epsilon_i^b})$\;
       $\alpha_0^c \longleftarrow \middle(\gamma(\hat{a}) \cup \gamma(\hat{b})) - \sum_{i=1}^{n} \alpha_i^c\middle(\vect{\epsilon_i^a} \cup \vect{\epsilon_i^b})$\;
       {\bf return} $(\alpha_0^c, \alpha_1^c, \dots{}, \alpha_n^c, \beta^c)$ 	\tcc*{$\scriptstyle{\bf MUB}$}
     }
}
$\beta^c \longleftarrow \dev(\gamma(\hat{a}) \cup \gamma(\hat{b}))$, $\alpha_0^c \longleftarrow \middle(\gamma(\hat{a}) \cup \gamma(\hat{b}))$, {\bf return} $(\alpha_0^c, \beta^c)$ \tcc*{$\scriptstyle{\bf UB}$}		\;
  \caption{Join of two constrained affine forms}\label{algo:mub_itv_min}
\end{algorithm}\DecMargin{1em}

\begin{example}
To complete the analysis of the running example of Section~\ref{introduction}, 
the join of the abstract values for $y$ on the two branches must be computed:
$$
\left\lbrace \begin{array}{l} \Phi^a = 1\times[-1,1]\times[-1,1]\times [-1,1] \\
\hat{a} = 0.5+0.5\varepsilon_1 \\
\gamma(\hat{a}) = [0,1]
\end{array} \right.
\left\lbrace \begin{array}{l} \Phi^b =  1\times[-1,-0.444]\times[-1,1]\times[-1,1] \\ 
\hat{b} =  14.93395+13.9\varepsilon_1+0.96605\varepsilon_3 \\
\gamma(\hat{b}) = [0.0679,9.7284]
\end{array} \right.
$$
$\hat{a}$ and $\hat{b}$ are in generic positions, and so are $\varepsilon_1^a$ and $\varepsilon_1^b$,
but condition $\middle(\vect{\epsilon_1^b}) \geq \middle(\vect{\epsilon_1^a} \cup \vect{\epsilon_1^b})$ 
is not satisfied, so that the join gives the following minimal upper bound:
$$
\left\lbrace\begin{array}{l}
\Phi^c = 1\times[-1,1]\times[-1,1]\times[-1,1] \\
\hat{c} = 4.8642+4.8642\eta_1, \; \gamma(\hat{c}) = [0,9.7284]
\end{array} \right.
$$
\end{example}

\begin{example}
Let us now consider a second example:
 $$
\left\lbrace \begin{array}{l} \Phi^a = 1\times[-1,0]\times[-1,1]\\
\hat{a} = 1+2\varepsilon_1-\varepsilon_2, \; \gamma(\hat{a}) = [-2,2]
\end{array} \right.
\left\lbrace \begin{array}{l} \Phi^b =  1\times[-1,1]\times[0,0.5] \\ 
\hat{b} =  4+3\varepsilon_1- \varepsilon_2, \;\gamma(\hat{b}) = [-2,7]
\end{array} \right.
$$
$\hat{a}$ and $\hat{b}$ are in generic positions, as well as $\varepsilon_1^a$ and $\varepsilon_1^b$, while
$\varepsilon_2^a$ and $\varepsilon_2^b$ are not; the join gives the following minimal upper bound:
$$
\left\lbrace\begin{array}{l}
\Phi^c = 1\times[-1,1]\times[-1,1]\times[-1,1] \\
\hat{c} = \frac{5}{2}+2\varepsilon_1+\frac{5}{2}\eta_1, \; \gamma(\hat{c}) = [-2,7]
\end{array} \right.
$$
\end{example}

\subsection{Join operator in the general case}
\label{generalmubs}

As in the unconstrained case \cite{arxiv2}, mubs for the global order
on {\em constrained affine sets} are difficult to characterize. Instead of
doing so, we choose in this paper to describe a simple yet efficient way
of computing a good over-approximation of such mubs, relying on 
Algorithm 1 for mubs with minimal concretisation for {\em 
constrained affine forms}.

First, we need two new operators. 
To each constrained affine set $X=(C^X,P^X,\Phi^X)$,
where $(C^X,P^X) \in {\cal M}(n+1,p) \times {\cal M}(m,p)$, and for each $k=
1,\ldots,p$, we associate the constrained affine form 
$$!^k X=
\biggl(c_{0,k} + \sum_{i=1}^n c_{i,k}\epsilon_i
+\sum_{j=1}^m p_{j,k} \eta_j,\ 0,\ \Phi^X \biggr) \enspace .$$ 
Basically, $!^k X$ takes
the $k$th column of $X$ (i.e. considers only the $k$th variable of the
environment) and treats the perturbation symbols $\eta_j$ as 
central symbols (like the $\epsilon_i$). Conversely, to each 
set of $p$ constrained affine forms $\hat{A}=(\hat{a_k})_k$ with 
$\hat{a_k}=\left(\alpha_{0,k} + \sum_{i=1}^n \alpha_{i,k} \epsilon_i, \ \beta_k, \Phi \right)$ (with the
same $\Phi$, and $k=1, \ldots, p$), we associate the constrained affine form
$?^l \hat{A}=(C,P,\Phi)$ with $(C,P) \in {\cal M}(l+1,p)\times {\cal M}(n-l+p,p)$
and $c_{i,k}=\alpha_{i,k}$, $i=0,\ldots,l$ and $k=1,\ldots,p$; 
$p_{j-l,k}=\alpha_{j-l,k}$, $j=l+1,\ldots,n$ and $k=1,\ldots,p$; 
$p_{n-l+k,k}=\beta_k$, $k=1,\ldots,p$, the rest being equal to zero. 

This operator basically embeds a set of $p$ constrained affine forms with
the same constraint on the noise symbols, into a constraint affine set such
that symbols $\epsilon_i$ up to $l$ are considered as central noise symbols,
whereas further symbols are considered as perturbation noise symbols; and
the perturbation term of each constrained affine form is considered to create
a new independent perturbation noise symbols (hence creating $p$ new
perturbation noise symbols in $?^l \hat{A}$).
 
\begin{definition}
\label{mub-def-gen}
Let $X=(C^X,P^X,\Phi^X)$ and $Y=(C^Y,P^Y,\Phi^Y)$ be two constrained
affine sets with $(C^X,P^X)$ and
$(C^Y,P^Y)$ in ${\cal M}(n+1,p) \times {\cal M}(m,p)$. 
We define 
$$J(X,Y)=?^n \left(!^k(C^X,P^X,\Phi^X \cup \Phi^Y)
\vee !^k(C^Y,P^Y,\Phi^X \cup \Phi^Y)\right)_{1 \leq k \leq p}$$ 
the join of $X$ and $Y$. 
It defines an upper bound of $X$ and $Y$.
\end{definition}

\begin{example}
Consider, for all noise symbols in $[-1,1]$, constrained affine sets X defined by $x_1=1 + \varepsilon_1, \; x_2=1 + \varepsilon_2$,
and $y_1=1 + \eta_1, \; y_2=1 + \eta_1$. Considering first the 1D cases, we have $x_1 \leq y_1$ and $x_2 \leq y_2$. However we do not 
have $X \leq Y$ for the global order of Definition \ref{constrainedpreorder}, but we have  $X \leq Z$ with Z defined by $z_1=1+\eta_1$ 
and $z_2=1+\eta_2$, constructed with the $?$ operator.
\end{example}

\section{Experiments}
\label{experiments}

In this section, we compare results\footnote{sources of the
examples are available online \url{http://www.lix.polytechnique.fr/Labo/Khalil.Ghorbal/CAV2010}} 
we obtain with our new domain, called constrained T1+, in its APRON implementation, with
the octagon and polyhedron APRON domains, the unconstrained T1+\cite{CAV09}, and the result  
given by our FLUCTUAT analyzer on the real value of variables. FLUCTUAT implementing a reduced
product of affine sets with intervals, this comparison demonstrates the interest of the logical 
product approach, with respect to a classical product. 
Our constrained T1+ implementation allows to choose as a parameter of the analysis, the APRON domain 
we want to use to abstract the constraints on noise symbols. However, at this stage, conditionals are interpreted 
only for the interval domain, we thus present results for this domain only.

Table~\ref{benchs} shows the numerical range of a variable of interest of each test case and for each domain, after 
giving the exact range we would hope to find.
\begin{table}
\vspace{-.6cm}
\caption{Comparison of Constrained T1+ with APRON's abstract domains \label{benchs}}
\begin{tabular}{|l|c|c|c|c|c|c|}
\hline
		& Exact	& Octagons 	& Polyhedra 	& T1+ 	& Fluctuat	& Constr. T1+	\\
\hline
InterQ1		& $[0,1875]$		& $[-3750,6093]$ &	$[-2578,4687]$ & $[-2e6,5e6]$ & $[0,2500]$ & $[-312,1875]$	\\
\hline
Cosine		& $[-1,1]$	& $[-1.50, 1.0]$	& $[-1.50, 1.0]$	& $[-4.43,4.71]$	& $[-1.073,1]$	& $[-1,1]$	\\
\hline
ItvPoly	& $x\geq3$ 	& $x\geq-2.21$ 	& $x\geq-2$ & top	& $x\geq-2$ & $x\geq-2$ 	\\
\hline
InterL2	& $\{0.1\}$	& $[-1,1]$	& $[0.1,0.4]$	& $[-1,1]$ & $[-1,1]$ & $[0.1,1]$		\\
\hline
InterQ2	& $\{0.36\}$	& $[-1,1]$	& $[-0.8,1]$	& $[-1,1]$ & $[-1,1]$ & $[-0.8,1]$		\\
\hline
\end{tabular}
\vspace{-.6cm}
\end{table}
\verb!InterQ1! combines linear tests with quadratic expressions, only constrained T1+ finds the right upper bound of the invariant.
\verb!Cosine! is a piecewise 3rd order polynomial interpolation of the cosine function: once again, 
only constrained T1+ finds the exact invariant.
\verb!InterL2! (resp. \verb!InterQ2!) computes a piecewise affine (resp. quadratic) function of the input, 
then focuses on the inverse image of $1$ by this function.
Note that our domain scales up well, as was the case with Taylor1+ (see \cite{CAV09} for benchmarks) 
while giving results that are often better than all the other domains presented here. As a matter of fact, 
for an interval domain for the noise symbols, all abstract transfer functions are linear or at worst quadratic in the
number of noise symbols appearing in the affine forms. 
Moreover, the superiority of the logical product approach over the reduced product (FLUCTUAT) is clearly demonstrated.

\section{Conclusion, and future work}
\label{futurework}

In this paper, we studied  the logical product of the domain of affine
sets with sub-polyhedric domains on noise symbols, 
although the framework as described here is much more general. We concentrated on such abstract domains 
for $\cal A$ for practical reasons, in order to have actual algorithms to compute 
the abstract transfer functions. %Let us mention

However, in some embedded control systems, %, such as the one described in \cite{DASIA09}, 
quadratic constraints appear already on the set of
initial values to be treated by the control program, or as a necessary
condition for behaving well, numerically speaking. For example in \cite{DASIA09}, as in a large class
of navigation systems, the control program manipulates normalized quaternions, that describe the current 
position in 3D, of an aircraft, missile, rocket etc. We think that a combination of zonotopes with 
quadratic templates \cite{ESOP2010} in the lines of this article would be of interest to analyze these programs. 
\comment{Hence we need to be able to represent, in the invariants of such
systems, constraints of the form:
$$q^2_1+q^2_2+q^2_3+q^2_4=1$$
where $q=(q_1,q_2,q_3,q_4) \in \R^4$ is a quaternion.

\paragraph{Constraints in maxplus polyhedra}
}

Also, as noticed in \cite{SAS2008}, maxplus polyhedra encompass a large subclass
of disjunctions of zones; hence, by combining it with affine
sets, we get another rather inexpensive way to derive a partially
disjunctive analysis from affine forms (another with respect to the ideas
presented in \cite{arxiv1}).

Another
future line of work is to combine the ideas of this paper with the ones
of \cite{SAS2007} to get better under-approximation methods in static
analysis.

\bibliographystyle{plain}
\bibliography{biblio}

\end{document}